\documentstyle[11pt]{article}

\def\be{\begin{equation}}
\def\ee{\end{equation}}
\def\ba{\begin{eqnarray}}
\def\ea{\end{eqnarray}}
\def\v#1{\vert #1 \rangle}

\newcommand{\R}{\mbox{I \hspace{-0.82em} R}}

\newcommand{\x}{{\bf x}}
\newcommand{\p}{{\bf p}}

\newcommand{\etab}{{\bar{\eta}}}
\newcommand{\detab}{{\partial_{\etab}}}

\newcommand{\sn}{\smallskip\newline}

\newcommand{\bns}{\bigskip\newline$\mbox{\qquad}$}
\newcommand{\mbo}{{\mbox{ }}}

\def\A{{\cal{A}}}
\def\M{{\cal{M}}}
\begin{document}

\title{ {\Large{\bf ON PATH INTEGRATION ON NONCOMMUTATIVE 
GEOMETRIES\thanks{Based on two talks given in the
\it Minisemester on Quantum Groups and Quantum Spaces, \rm \newline
$\mbox{ \quad }$ held Nov.95 at the Banach Center, Warsaw, Poland}}}\\ }

\author{A{\,}C{\,}H{\,}I{\,}M K{\,}E{\,}M{\,}P{\,}F{\,}\\ \\ \it
Department of Applied Mathematics \& Theoretical Physics\\
\it and Corpus Christi College in the University of Cambridge\\
\it Silver Street, Cambridge CB3 9EW, U.K.\\
\it  E-mail: a.kempf@amtp.cam.ac.uk \rm}

\date{ }
\maketitle

\vskip-8.7cm
{\tt $\mbo$ \hskip9.5cm DAMTP/96-10}\newline
{\tt $\mbo$ \hskip10.12cm  hep-th/9603115}\rm
\vskip8.4cm

{\small {\bf Abstract.} We discuss a recent approach 
to quantum field theoretical path integration
on noncommutative geometries which imply UV/IR regularising finite 
minimal uncertainties in positions and/or momenta. One class of
such noncommutative geometries arise as
`momentum spaces' over curved spaces, for which
we can now give the full set of commutation relations in
coordinate free form, based on the Synge world function.
\bns
{\bf 1. Introduction.} 
A crucial example of noncommutative geometry \cite{connes}
is the quantum mechanical
phase space with its noncommuting `coordinate functions' $\x_i$ and
$\p_j$. We investigate the possibility that also the
position and momentum spaces acquire noncommutative geometric features, 
i.e. we consider associative 
Heisenberg algebras $\A$ generated by elements $\x_i,\p_j$, now allowing
\be
[\x_i,\x_j] \ne 0, \qquad \quad [\p_i,\p_j] \ne 0
\label{ekam}
\ee 
and also:
\be
[\x_i,\p_j] = i\hbar\mbo ( \delta_{ij} + \alpha_{ijkl} \x_k\x_l +
\beta_{ijkl} \p_k\p_l + ... )
\label{dekam}
\ee
We restrict ourselves to relations that allow the involution
$\x^*_i = \x_i, \p^*_i = \p_i$, i.e. for which `$^*$'
extends to an anti algebra homomorphism. To motivate the 
particular form of relation Eq.\ref{dekam}, let this relation 
be represented on a dense
domain $D \subset H$ in a Hilbert space $H$, i.e.
both the $\x_i$ and the $\p_j$ are to be represented as symmetric 
operators on $D$. Assuming, e.g. in the simplest case of one dimension 
$\alpha,\beta>0$ and 
$\alpha \beta < 1/\hbar^2$, together with the usual definition
of uncertainties
\be
(\Delta x)^2_{\v{\psi}} :=
\langle \psi \vert (\x - \langle \psi \vert 
\x \vert \psi \rangle)^2 \v{\psi } 
\ee
yields
\begin{equation}
\Delta x \Delta p \ge \frac{\hbar}{2} \left( 1 + \alpha (\Delta x)^2 
+ \alpha \langle \x\rangle ^2
+ \beta (\Delta p)^2 + \beta \langle \p\rangle ^2 \right)
\label{ucr}
\end{equation}
As is not difficult to check Eq.\ref{ucr} implies that there are
finite minimal uncertainties $\Delta x_0 = (1/\beta\hbar^2 -\alpha)^{-1/2}$
and $\Delta p_0 = (1/\alpha\hbar^2 -\beta)^{-1/2}$,
so that there appears a `minimal uncertainty gap' (all $\v{\psi}$ normalised):
\be
\forall \mbo \v{\psi} \in D: \quad \Delta x_{\v{\psi}} \ge \Delta x_0
\mbox{ \qquad and \qquad } \Delta p_{\v{\psi}} \ge \Delta p_0
\label{gap}
\ee
Physically, since $\alpha$ and $\beta$ can be assumed small, we have
ordinary quantum mechanical behaviour on medium scales. 
The presence of a finite $\Delta x_0$, physically relevant
in the ultraviolet, e.g. at the Planck scale, 
can be motivated from studies in string theory and
quantum gravity, see e.g. \cite{townsend}-\cite{garay}. The presence
of a finite $\Delta p_0$, relevant in the infrared, i.e. on large scales,
may be motivated from the absence of plane waves (i.e. of sharp 
localisations in momentum space) on generic curved spaces, see \cite{ak-np}.

It should be interesting to apply A.Connes programme to such
generalised phase spaces, i.e. in particular to develop the
corresponding differential and integral calculus, though we will here
not follow this approach. Instead we will focus on the field theoretic
and gravity-related aspects of the ansatz.
\bns
{\bf 2. Representation Theory.} 
Due to the minimal uncertainty gap Eq.\ref{gap},
algebras $\A$ which imply minimal uncertainties cannot find spectral 
representations of neither the $\x$ nor $\p$
(recall that e.g. $\x \v{\psi} = x \v{\psi}$ implies 
$\Delta x_{\v{\psi}} = 0$). In fact we give up (essential) self-adjointness
of the $\x$ and $\p$ on physical domains $D$
to retain only their symmetry. 
While giving up self-adjointness
is necessary for the description of the new short distance
behaviour, the symmetry is sufficient for the definition of
uncertainties to be well defined, and to ensure the realness of
expectation values, e.g. $\forall \mbo \v{\psi} \in D: \mbo 
\langle \psi \vert \x \vert \psi \rangle \in \R$.
The situation has been studied in explicit representations. The
deficiency indices have, e.g. in one dimension,
been found as $(1,1)$ and the self-adjoint
extensions and eigenvectors of the $\x$ and $\p$ have been calculated.
The spectra of the $\x$ and $\p$ are discrete, with the eigenvectors 
in $H$, though of course not in $D$.
Unlike the case of the ordinary commutation relations 
there are now no sequences of
physical states which would approximate point localisation 
i.e. $\exists\!\!\!/ \mbo\{ \v{\psi_n} \in D\}: \mbo 
\lim_{n\rightarrow \infty} (\Delta x)_{\vert \psi_n \rangle} = 0$
and analogously for the momentum.
\sn
We give two examples of Hilbert space
representations for the one-dimensional case.
With $L,K$ length and momentum scales obeying $LK=(q^2+1)/4$ and
$q>1$ the relation
\be
[\x,\p] = i\hbar 
\left(1 + (q^2-1)\left( \x^2/4L^2 + \p^2/4K^2\right)\right)
\label{crkl}
\ee
can be represented e.g. on polynomials in $\etab$ through 
\be
\x.\psi(\etab) = L (\etab + \detab) \psi(\etab)
\qquad 
\p.\psi(\etab) = iK (\etab - \detab) \psi(\etab)
\label{bf1}
\ee
where the differentiations are to be evaluated algebraically by
commuting $\detab$ to the right using the `Leibnitz rule'
\be
\detab \etab - q^2 \etab \detab = 1
\label{ed}
\ee
and where the scalar product now reads $\langle \psi_1\vert \psi_2\rangle
= \psi_1^*(\detab)\psi_2(\etab)\vert_{\etab=0}$. 
In the special case $\alpha = 0$ the algebra $\A$ 
still finds spectral representations of $\p$. A convenient possibility is
\be
\p.\psi(p) = p\psi(p), \qquad \x.\psi(p) = i\hbar(1+\beta \p^2)\partial_p
\psi(p)
\label{mome}
\ee
with the scalar product $\langle \psi_1\vert \psi_2\rangle
= \int_{-\infty}^{\infty} dp \mbo (1+\beta p^2)^{-1} \psi_1^*(p) \psi_2(p)$.
Using the Bargmann Fock representation Eqs.\ref{bf1},\ref{ed}
physical features such as the
scalar product of maximal localisation states have been calculated 
\cite{ak-hh-1} and these results nicely reduce to the results previously
obtained working in the momentum representation Eq.\ref{mome} 
in the special case $\alpha =0$, see \cite{ak-gm-rm-prd}.
This is however nontrivial, and for generic algebras $\A$ not all
Hilbert space representations will be unitarily equivalent. 
We conjecture that
the unitary equivalence does hold in the sense in which it holds for
the ordinary commutation relations, for those cases considered 
so far, i.e. with positive (matrices) $\alpha,\beta$.
For  $\alpha$ or $\beta$ negative nonequivalent representations can be 
found \cite{ak-gm-2}.
\sn
To the best of my knowledge relations equivalent to the 
$\etab,\detab$-relations Eq.\ref{ed} were first studied 
in \cite{arik}, and the $n$- dimensional $SU_q(n)$- covariant
generalisation was first studied in \cite{woro}. 
The relation Eq.\ref{crkl}, its physical implications and 
that one of its Hilbert space representations is on
the algebra of these $\etab,\detab$ was however not known, see
also \cite{ak-lmp-bf,ak-jmp-bf,klimyk}.
That the generalised Heisenberg algebra of positions and momenta
with commutation relations Eq.\ref{crkl} (and generalisations)
imply finite minimal uncertainties $\Delta x_0,\Delta p_0$, and the
crucial interplay between the functional analysis of the $\x$ and $\p$
was first shown in \cite{ixtapa,ak-jmp-ucr}. 
The special case $\alpha =0$ was first studied in \cite{ak-gm-rm-prd}
and the there obtained results on maximal localisation states have been 
extended to the general situation in \cite{ak-hh-1}. Our
approach to quantum field theory on noncommutative 
geometries, that we will now describe, was first studied 
in \cite{ak-np,prag}.
\bns
{\bf 3. Path Integration.} Let us illustrate the ansatz with
the simple example of charged euclidean $\phi^4$-theory. Its
partition function
\be
Z[J] := N \int D\phi\mbo e^{\int d^4x\mbo
\phi^* (\partial_i\partial_i - \mu^2)\phi 
 - \frac{\lambda}{4!}(\phi \phi)^*\phi \phi + \phi^*J+J^*\phi}
\label{pfx}
\ee
we rewrite as 
\be
Z[J] = N \int_D D\phi\mbo e^{-\bf tr\rm\left(\frac{l^2}{\hbar^2} 
(\p^2+m^2c^2).\vert\phi\rangle\langle\phi\vert + \frac{\lambda l^4}{4!}
\vert\phi *\phi\rangle\langle\phi *\phi\vert + \vert\phi\rangle\langle J\vert
+\vert J\rangle\langle\phi\vert\right)}
\label{pf}
\ee
where, to make the units transparent, 
we introduced an arbitrary positive length to render
the fields unitless ($l$ could trivially be reabsorbed in the fields).
\sn
We recover Eq.\ref{pfx} from Eq.\ref{pf} by assuming the ordinary
relations $[\x_i,\p_j]= i \hbar \delta_{ij}$ in $\A$ and by choosing 
the spectral representation of the $\x_i$ where with 
$\phi(x) := \langle x\vert \phi\rangle$
\be
\x_i.\phi(x) = x_i\phi(x), \qquad \p_i.\phi(x) = -i\hbar\partial_{x_i}\phi(x)
\ee
and where
\be
\bf tr\rm(q) = \int d^4x \mbo \langle x\vert q\vert x\rangle
, \qquad (\phi_1 * \phi_2)(x) = \phi_1(x) \phi_2(x)
\ee
The pointwise multiplication $*$, used to express point interaction, 
is (and can also on noncommutative geometries be kept) 
commutative for bosons.
Since fields are in a representation of $\A$, similar to quantum mechanical 
states, we formally extended Dirac's 
bra-ket notation for states to fields. In Eq.\ref{pf}
this yields a convenient notation for the functional analytic structure 
of the action functional, but of course, the quantum mechanical
interpretation does not simply extend, see e.g. \cite{fdl}.
The space $D$ of fields that is formally to be summed over can be taken
to be the dense domain $S_{\infty}$ in the Hilbert space $H$ 
of square integrable fields.                 
\sn
Generally, the unitary transformations that map from one
Hilbert basis to another have trivial determinant, so that no anomalies are
introduced into the field theory and changes of basis can be performed arbitrarily,
in the action functional, in the Feynman rules or in the end results
of the calculation of $n$- point functions.
Choosing any other, e.g. discrete, Hilbert basis $\{\v{n}\}$ in $D$ we
have $\phi_n = \langle n\vert \phi \rangle$ so that
\be
(\p^2 +m^2)_{nm} = \langle n \vert \p^2 + m^2 \vert m\rangle
\label{e1}
\ee
and
\be
* = \sum_{n_i} L_{n_1,n_2,n_3} \vert n_1\rangle \otimes
\langle n_2\vert \otimes \langle n_3 \vert 
\ee
Thus:
\be
Z[J] = N \int_D D \phi \mbo e^{ -\frac{l^2}{\hbar^2}\mbo
\phi_{n_1}^* (\p^2 +m^2)_{n_1n_2} \phi_{n_2}
-\frac{\lambda l^4}{4!} L^*_{n_1n_2n_3}
L_{n_1n_4n_5} \phi^*_{n_2} \phi^*_{n_3} \phi_{n_4}\phi_{n_5}
+ \phi^*_n J_n + J^*_n \phi_n }
\label{d1}
\ee
Pulling the interaction term in front of the path integral,
completing the squares, and carrying out the gaussian integrals
yields 
\be
Z[J] = N' e^{-\frac{\lambda l^4}{4!} L^*_{n_1n_2n_3} L_{n_1n_4n_5}
\frac{\partial}{\partial J_{n_2}}
\frac{\partial}{\partial J_{n_3}}
\frac{\partial}{\partial J^*_{n_4}}
\frac{\partial}{\partial J^*_{n_5}}}
\mbo
e^{-\frac{\hbar^2}{\l^2} J^*_n (\p^2 +m^2)^{-1}_{nm} J_m}
\ee
Generally, the inverse of $(\p^2 +m^2)$ on $D$ may not be unique in which
case we choose a self-adjoint extension in which it can be 
diagonalised and inverted, i.e. physically one chooses
boundary conditions.
We obtain the Feynman rules:
\be
\Delta_{n_1n_2} = \left(\frac{-\hbar^2}{l^2(\p^2
+m^2)}\right)_{n_1n_2}
\qquad
\mbox{and } \qquad
\Gamma_{n_1n_2n_3n_4} = -\frac{\lambda l^4}{4!}
L^*_{mn_1n_2} L_{mn_3n_4} 
\label{d2}
\ee
Let us recall that the usual formulation
of partition functions, such as e.g. Eq.\ref{pfx}, implies that 
$\p^2$ can be represented as the d'Alembertian on a spectral representation 
of the $\x_i$, so that $\p_i$ is represented
as $-i\hbar\partial_i$, i.e. it is implied 
that $ [\x_i,\p_j] = i\hbar \delta_{ij}$.
It is crucial that in our formulation of 
partition functions in abstract form, such as in
Eq.\ref{pf}, the commutation
relations of the underlying algebra $\A$ are not implicitly fixed
and can be generalised, e.g. to the form of 
Eqs.\ref{ekam},\ref{dekam}.

Representing $\A$ on some $D$ in a Hilbert space $H$
with an e.g. discrete Hilbert basis $\{\v{n}\}$ (the 
Hilbert space is separable), one straightforwardly obtains the
Feynman rules through Eqs.\ref{d1}-\ref{d2}. The formalism thus allows
for example to explicitly check 
noncommutative geometries on  
UV and IR regularisation. So far, regularisation has been shown for 
certain examples of geometries $\A$ that imply
minimal uncertainties $\Delta x_0, \Delta p_0$, see \cite{ak-np,prag},
and very recently \cite{nr,reg}.

Without going into details, 
let us only remark that in this context the structure of the pointwise
multiplication $*$ that describes local
interaction is crucial. Due to the absence of a position representation,
it is nonunique in the case of $\Delta x_0 > 0$, though
there are still `quasi-position representations' \cite{ak-gm-rm-prd,ak-hh-1}, 
built on maximal localisation states, which can be
used to establish the locality and causality properties 
of pointwise multiplications. Generally in our approach an
interaction is
observationally local if any formal nonlocality of $*$ is not larger than
the scale of the nonlocality $\Delta x_0$ inherent in the 
underlying space. 
Thus, a UV-regular, formally slightly nonlocal $*$, 
can describe an observationally strictly local interaction. Further
studies on locality and regularisation are in progress \cite{ak-gm-2}. 
An alternative approach, based on the canonical formulation 
of field theory is \cite{doplicher1}.
\bns
{\bf 3. Curvature and Noncommmutativity.} 
Assuming for $\A$ commutation 
relations of the type of Eqs.\ref{ekam},\ref{dekam} means in particular that
the momenta are no longer the generators of what would be 
infinitesimal translations
on flat space, because then $\x_i \rightarrow \x_i^{\prime} :=
\x_i -\alpha_j  [\x_i,\p_j] /i\hbar \ne \x_i - \alpha_i$.
We intend to show that certain commutation relations of the 
type of Eqs.\ref{ekam},\ref{dekam} can arise for momenta on curved space.

Let $\M$ be a (pseudo-) Riemannian manifold where we for simplicity
assume vanishing torsion.
Due to the path dependence of parallel transport
any definition of 4-{\it vectors} of positions $\x$ 
and momenta $\p$ must relate
to measurement prescriptions that specify some point $Q$ on the
manifold for $\x$ and $\p$ to live in the (co-) tangent spaces to $Q$.
Choosing an event $Q \in \M$ there exist
normal (i.e. geodesically convex) neighbourhoods $U(Q)$ in which
the Synge world function $\sigma$, see e.g. \cite{dewitt,fulling} 
is well defined through
\be
\sigma(Q,x) := \frac{1}{2} s^2
\ee
with $s$ being the geodesic distance between events $Q$ and $x$.
Covariant differentiation with respect to $Q$
\be
\sigma^\mu(Q,x) := -g^{\mu\nu}(Q)\mbo \nabla_{\nu} \sigma(Q,x)
\ee
yields the vector field of geodesic coordinates of the event $x$, i.e.
$\sigma^{\mu}(Q,x)\in T_Q(\M)$ is the initial tangent vector of
the geodesic that reaches $x$ as its parameter reaches 1.
\sn
We now define for each $Q$ a local Heisenberg algebra $\A_Q$, generated by
elements $\x_Q^\mu,\p_{Q\nu}$ which, as vectors, lie in the (co-) tangent
spaces to $Q$.
We start by defining $[\x_Q^{\mu},\x_Q^{\nu}]=0$
and by mapping the spectrum of 
the $\x_Q^{\mu}$ onto the manifold through the exponential map. Hence the
$\x_Q^\mu$ correspond to the measurement of the coordinates 
$\sigma^\mu$ of the event $x$ in the geodesic coordinate system, or 
\it frame, \rm with origin $Q$.
The momentum operators now be defined to generate the change of the geodesic
coordinates as we move the origin of the geodesic 
frame from the event $Q$, by
$\alpha\in T_Q(\M)$, to the infinitesimally distant $Q^{\prime}$
\be
\x_Q^\mu \rightarrow \x_{Q^{\prime}}^\mu = \x_Q^\mu -
\alpha^\nu \mbo \sigma^\mu{}_\nu(Q,\x_Q) =: \x_Q^\mu - \frac{1}{i\hbar} 
\alpha^{\nu} [\x_Q^{\mu},\p_{Q\nu}] 
\label{trfr}
\ee
with the usual definition $\sigma^\mu{}_\nu := \sigma^\mu{}_{;\nu}$. 
We read off the required commutation relations:
\be
[\x_Q^{\mu},\p_{Q\nu}] = i\hbar \mbo \sigma^{\mu}{}_{\nu}(Q,\x_Q)
\label{xpg}
\ee
The Jacobi identities and $\p_\nu=\p_\nu^*$ then yield
\be
[\p_{Q\mu}, \p_{Q\nu}] = i \hbar
\{\p_{Q\alpha}, \sigma_\rho^{-1}{}^\alpha \sigma^\epsilon{}_{[\nu}
\sigma^{\rho}{}_{\mu],\epsilon} \}
\label{ppcr}
\ee
The Heisenberg algebras $\A_Q$ are a special 
case of Eqs.\ref{ekam},\ref{dekam}
so that the path integral in its abstract formulation Eq.\ref{pf} can 
be evaluated through Eqs.\ref{e1}-\ref{d2}, as discussed.
There exist spectral representations of the $\x_Q$
with $\p_{Q\mu}. \psi(x) = - i \hbar \left( 1/2 \sigma^\epsilon{}_{\mu,\sigma}
+ \sigma^\epsilon{}_\mu \partial_{x^\epsilon}\right) \psi(x)$. 
Studies into whether $Q$-independence of the 
action functional can be understood as a gauge principle are 
in progress.

A detailed comparison of the approach with the conventional treatment
is in preparation, and we mention here only a few general points.
Path integrals such as Eq.\ref{pfx} are conventionally 
adapted to the curved space
in the position representation, by 
replacing ordinary derivatives by covariant derivatives and by 
adapting the measure introducing
$\sqrt{\vert -g\vert}$, see e.g. \cite{birrel}. This is in
complete analogy to how observable fields would have to be treated.
Let us recall however, that for quantum fields this procedure, 
which singles out position space, spoils the beautiful 
quantum theoretical representation independence of the action 
functional. There then no longer exists an algebra $\A$ of positions and
momenta, for which one could freely choose a Hilbert space representation.
In our approach, we do still have a Heisenberg algebra of positions
and momenta, so that the abstract functional analytical structure Eq.\ref{pf}
of the path integral is preserved, and representations of $\A$ need not be
position representations, but could as well be a representation e.g. on a 
Hilbert space of orthogonal polynomials. 
Note that choosing the position representation the $\p_i$ do not act by 
simple covariant 
differentiation but by making use of the canonical tensor fields
that arise as covariant derivatives $\sigma^\mu{}_\nu$ 
of the world function $\sigma$.
\sn
We give the first correction terms explicitly.
It is possible to covariantly `Taylor expand' $\sigma^\mu{}_\nu$ within
some `radius' in $U(Q)$, see \cite{fulling}. 
To this end $\sigma^\mu{}_\nu$ and its higher
covariant derivatives are needed in the 
coincidence limit $Q=x$. 
As is well known, in the coincidence limit 
$\sigma^\mu{}_\nu(Q,Q)=\delta^\mu_\nu$, (i.e. also 
$\sigma_{\mu\nu}(Q,Q)=g_{\mu\nu}(Q)$). 
The next order term is the torsion tensor
$\sigma^{\mu}{}_{\nu\rho}(Q,Q) = -1/2 T^\mu{}_{\nu\rho}(Q)$ which,
if we had not taken it to vanish, would contribute a term linear in $\x$
on the RHS of the 
commutation relation Eq.\ref{xpg}. To the second order we have
$\sigma^{\mu}{}_{\nu\rho\tau}(Q,Q) = -1/3 (R^\mu{}_{\rho\nu\tau}(Q)
+ R^\mu{}_{\tau\nu\rho}(Q)) = -2/3 J^\mu{}_{\nu\rho\tau}(Q)$, see
\cite{fulling}, where the Jacobi curvature tensor is 
${J^{\mu}}_{\nu\alpha\beta} = \frac{1}{2}
 ({R^{\mu}}_{\alpha\nu\beta}
+ {R^{\mu}}_{\beta\nu\alpha})$, see \cite{mtw}.
The covariant Taylor expansion for $\sigma^\mu{}_\nu$ then yields
\be
[\x_Q^{\mu},\p_{Q\nu}] = i\hbar \mbo \delta_{\nu}^{\mu} - \frac{i\hbar}{3}
\mbo {J^{\mu}}_{\nu\alpha\beta}(Q) \mbo \x_Q^{\alpha} \x_Q^{\beta} 
+ ...
\label{xpge}
\ee
and
\be
[\p_{Q\mu},\p_{Q\nu}] = 
\frac{1}{2}\mbo i\hbar \mbo R^\rho{}_{\alpha\nu\mu}(Q)
\{\x_Q^\alpha,\p_{Q\rho}\} + ...
\label{ppge}
\ee
The higher terms in the covariant expansion of $\sigma^\mu{}_\nu$
are expressions
in covariant derivatives of $R$, which have been calculated by computer
to a few orders, \cite{fulling}.
\sn
It is instructive to derive the correction terms also by
the pedestrian method, i.e. in coordinates. 
\sn
Denote by $G_Q$ the geodesic coordinate system, or frame, with the
event $Q$ at it's origin.
Further, let the event $Q^{\prime}$ which has the coordinates $\alpha^{\mu}$
in $G_Q$ define the origin of a second geodesic frame $G_{Q^{\prime}}$.
The axes of $G_{Q^{\prime}}$ be starting off parallel to
those of the frame $G_Q$, i.e. the corresponding tangent vectors
are obtained through parallel transport along the geodesic connecting
$Q$ and $Q^{\prime}$.
\sn
Under the infinitesimal
translation of frames from $Q$ to $Q^{\prime}$
an event's coordinates $x_Q^{\mu}$ 
in $G_Q$ transform into it's coordinates $x_{Q^{\prime}}^\mu$ 
in $G_{Q^{\prime}}$ and vice versa, see Eq.\ref{trfr}.
The coordinates $x^\mu_{Q^\prime}$ of an event $x$ 
denote by definition the initial tangent vector of the geodesic 
$\gamma : [0,1] \rightarrow \M, d/ds\gamma(Q)= x^\mu_{Q^\prime},
\gamma(0) =Q, \gamma(1) =x$, where in $G_Q$: 
$\gamma : [0,1] \rightarrow \gamma^\mu(s) \in T_Q(\M)$.
The end point of this geodesic needs to be calculated in the
frame $G_Q$ to obtain the coordinates $x_Q$ of $x$.
The ordinary Taylor theorem yields:
\be
x_Q^\mu = e^{d/ds}\gamma^\mu(s)\vert_{s=0}
\ee
In $G_Q$ holds $\Gamma(Q)=0$ so that the geodesics through 
the event $Q^{\prime}$ obey
\be
\frac{d^2\gamma^{\mu}}{ds^2}(Q)+{\Gamma^{\mu}}_{\beta_1\beta_2,\nu}(Q) \mbo
\alpha^{\nu} \frac{d\gamma^{\beta_1}}{ds}(Q)\mbo
\frac{d\gamma^{\beta_2}}{ds}(Q) + {\cal{O}}(\alpha^2) = 0
\ee
and thus generally to first order in $\alpha$:
\be
\frac{d^{n+2}}{ds^{n+2}} \gamma^\mu(Q) = - 
\Gamma^\mu{}_{\beta_1\beta_2,\nu\rho_1...\rho_n}(Q)\mbo \alpha^\nu 
x_{Q^\prime}^{\beta_1} x_{Q^\prime}^{\beta_2}
x_{Q\prime}^{\rho_1} ... x_{Q\prime}^{\rho_n} + {\cal{O}}(\alpha^2)
\label{lchr}
\ee
To first nontrivial order the 
coordinates $x_Q^{\mu}$ in the frame $G_Q$ of the event with the 
coordinates $x_{Q^{\prime}}^\mu$ in the frame $G_{Q^{\prime}}$ therefore
read 
\be
x_Q^{\mu} = {x_{Q^{\prime}}^{\mu} + \alpha^{\mu}  - \frac{1}{2}\mbo
{\Gamma^{\mu}}_{\alpha\beta,\nu}(Q)\mbo \alpha^{\nu} 
x_{Q^{\prime}}^{\alpha} x_{Q^{\prime}}^{\beta} + 
{\cal{O}}(\alpha^2,x_{Q^\prime}^3})
\label{trafoo}
\ee
so that
\be
x_{Q^\prime}^\mu =  x_Q^{\mu} - \alpha^{\mu} + \frac{1}{2}\mbo
{\Gamma^{\mu}}_{\alpha\beta,\nu}(Q)\mbo \alpha^{\nu} 
x_Q^{\alpha} x_Q^{\beta} + {\cal{O}}(\alpha^2,x_Q^3)
\label{trafo}
\ee
Hence, with Eq.\ref{trfr}:
\be 
\sigma^{\mu}{}_{\nu}(Q,x) = \delta^{\mu}_{\nu} - \frac{1}{2}\mbo
{\Gamma^{\mu}}_{\alpha\beta,\nu}(Q)\mbo 
x^{\alpha} x^{\beta} + {\cal{O}}(x^3)
\ee
At the origin of geodesic coordinate systems, 
i.e. here at $Q$, holds \cite{mtw}:
$
{\Gamma^{\mu}}_{\alpha\beta,\nu}(Q) = -\frac{1}{3}
\left({R^{\mu}}_{\alpha\beta\nu}(Q) + {R^{\mu}}_{\beta\alpha\nu}(Q)\right)
$, 
so that we recover indeed the above results Eqs.\ref{xpge},\ref{ppge}.
We also proved  
that (at the origin of geodesic frames)
all $\Gamma^\mu{}_{\nu_1(\nu_2,\nu_3...\nu_r)}(Q)$
are tensors (which has been shown by other methods, 
inductively in coordinates, in \cite{yamashita}). 
Note also that from Eq.\ref{lchr} follows $\forall r \ge 2: 
\Gamma^\mu{}_{(\nu_1\nu_2,\nu_3...\nu_r)}(Q)=0$. 

Apart from the suggested physical interpretation, we hope that our approach 
could provide a useful
framework for expressing certain noncommutative geometries
as the `dual' or momentum space to a curved space, and vice versa,
thereby possibly making available tools of ordinary geometry for
the description of certain noncommutative geometries. 
Work in this direction is in progress. 
\bns
{\bf Acknowledgement.} It is my pleasure to thank the organisers 
for the very kind hospitality at the Banach Center.

\end{document}